\documentclass[12pt]{article}

\usepackage[nosort]{cite}
\usepackage{cite}
\usepackage{amsmath,pifont,bbding}
\usepackage{epsfig}
\usepackage{amsfonts}
\usepackage{amssymb}
\usepackage{multirow}
\usepackage{enumerate}
\usepackage{subfigure}
\usepackage{slashed}

\usepackage{color}

\usepackage{subfigure}

\usepackage{caption}

\newcommand{\be}{\begin{equation}}
\newcommand{\ee}{\end{equation}}
\newcommand{\bee}{\begin{equation*}}
\newcommand{\eee}{\end{equation*}}
\newcommand{\bea}{\begin{eqnarray}}
\newcommand{\eea}{\end{eqnarray}}
\newcommand{\bean}{\begin{eqnarray*}}
\newcommand{\eean}{\end{eqnarray*}}

\begin{document}

\setcounter{page}{0}
\thispagestyle{empty}

\begin{flushright}
\end{flushright}

\vskip 8pt

\begin{center}
{\bf \LARGE {Baryogenesis from Strong CP Violation 
\vskip 10pt
and  the QCD Axion  }}
\end{center}

\vskip 12pt

\begin{center}
 {\bf  G\'eraldine  Servant$^{a,b}$ }
 \end{center}

\vskip 14pt

\begin{center}
\centerline{$^{a}${\it ICREA, Instituci\'o Catalana de Recerca i Estudis Avancats, Barcelona, Spain}}
\centerline{$^{b}${\it IFAE, Universitat Aut\`onoma de Barcelona, 08193 Bellaterra, Barcelona}}

\vskip .3cm
\centerline{\tt gservant@ifae.es}
\end{center}

\vskip 10pt

\begin{abstract}
\vskip 3pt
\noindent

The strong CP-violating 
parameter is small today as indicated by constraints on the  neutron electric dipole moment. 
 In the early universe,
the QCD axion has not yet relaxed to its 
QCD-cancelling minimum and
Êit is natural to wonder whether this large CP violation could be responsible for baryogenesis.
We show that strong CP violation from the QCD axion can be responsible for the matter antimatter asymmetry of the universe in the context of  cold electroweak (EW) baryogenesis if the EW phase transition is delayed below the GeV scale. This can occur naturally if the Higgs couples to a ${\cal O}(100)$ GeV dilaton, as expected in some models
 where the Higgs is a pseudo-Nambu Goldstone boson of a new strongly interacting sector at the TeV scale.
The only new relevant ingredients beyond the Standard Model in our framework are the QCD axion and an EW scale dilaton-like scalar field. 
The existence of such a second scalar resonance with a mass and properties similar to the Higgs boson will soon be tested at the LHC. In this context, the QCD axion would not only solve the strong CP problem, but also the matter anti-matter asymmetry and  dark matter.

\end{abstract}

\newpage

\tableofcontents

\vskip 13pt

\section{Introduction}

Understanding the generation of the matter-antimatter asymmetry of the universe $\eta=\eta_{10} \times 10^{-10}$, where $\eta_{10}=6.047 \pm 0.074$  \cite{Ade:2013zuv}, is one of the key motivations for physics beyond the Standard Model (SM). The SM  fails to satisfy two of the three Sakharov conditions needed for baryogenesis. First, as the electroweak (EW) phase transition is not first-order \cite{Rummukainen:1998as}, there is no sufficient departure from equilibrium in the standard cosmological evolution. Second, CP violation in the SM appears to be too small to explain the value of $\eta$. There is no absolute proof that CP violation in the SM is insufficient  for baryogenesis to work. However, so far, all attempts  to predict the value of $\eta$ with SM CP violation have failed. In particular, 
CP violation from the Cabibbo-Kobayashi-Maskawa matrix has been shown to be too small to play any role in electroweak baryogenesis \cite{Gavela:1993ts,Konstandin:2003dx}.   Some more recent  attempts were made in the context of {\it cold} baryogenesis, but they remain inconclusive \cite{Tranberg:2009de,Brauner:2012gu}.

There are too leading candidates to explain $\eta$: leptogenesis from out-of-equilibrium decay of heavy right-handed neutrinos before the EW phase transition \cite{Davidson:2008bu}, and electroweak baryogenesis during a first-order EW phase transition \cite{Morrissey:2012db,Konstandin:2013caa}. 
\footnote{Among alternatives, there are recent proposals  in which an asymmetry primarily produced in a new physics sector is transmitted to the SM, and in which no new source of baryon nor lepton number violation beyond the SM are needed \cite{Servant:2013uwa,Davidson:2013psa}. They are typically not as minimal as the leptogenesis or EW baryogenesis, but are interesting new routes for baryogenesis.}

While a large number of possibilities for new sources of CP violation arise in minimal TeV scale extensions of the SM and have been considered for baryogenesis, it is natural to wonder whether  the CP non conserving term in the SM QCD lagrangian 
\be
{\cal L} = \bar{\Theta}\frac{\alpha_s}{8\pi}  G_{\mu\nu a} \tilde{G}^{\mu\nu}_a \ \ \ \ , \ \ \ \ {\bar{\Theta}}= {\Theta +\mbox{arg det} {\bf M}_q}
\label{eq:thetabar}
\ee
could have played a role for baryogenesis.
The CP-violating $\bar{\Theta}$ term  is constrained today to be smaller than $10^{-11}$ from the absence of a measurable electric dipole moment for the neutron \cite{Baker:2006ts}.
The
$\Theta$ parameter characterizes  the non-trivial nature of the QCD vacuum. Because chiral transformations change the $\Theta$ vacuum once we include weak interactions and the quark mass matrix, the only physical observable angle is 
$\bar{\Theta}={\Theta + \mbox{arg det} {\bf M}_q}$ where $\mbox{\bf M}_q$ is the quark mass matrix.
The QCD angle $\Theta$, which is required to solve the  $U(1)_A$ problem \cite{'tHooft:1986nc}, and $\mbox{arg det} {\bf M}_q$ have nothing to do which each other and there is no reason why they should be tuned such that $|\bar{\Theta}| < 10^{-11}$. This is the so-called strong CP problem \cite{Peccei:2006as}.

The QCD vacuum energy depends on $\bar{\Theta}$ and is minimized at $\bar{\Theta}=0$.
Therefore, the puzzle is solved if $\bar{\Theta}$ is promoted to a dynamical field which relaxes naturally to zero, as advocated by  Peccei-Quinn (PQ) \cite{Peccei:2006as}.
This solution  postulates a new global axial symmetry $U(1)_{PQ}$ spontaneously broken by a scalar field 
$\Phi=(f_a+\rho(x)) e^{ia(x)/f_a}/{\sqrt{2}}$, where the Goldstone boson $a(x)$ is the axion.
New  heavy colored quarks with coupling to $\Phi$ generate a $G\tilde{G}$ term:
\be
  \frac{\alpha_s}{8\pi}   \ \frac{a(x)}{f_a} \ G_{\mu\nu a} \tilde{G}^{\mu\nu}_a.
 \label{eq:axion}
\ee
The axion couples to gluons, mixes with pions and couples to photons. Its couplings are all suppressed by the factor $1/f_a$ while its mass today satisfies 
\be
m_a f_a = m_{\pi} f_{\pi}  \sqrt{m_u m_d}/(m_u+m_d). 
\label{eq:axionmass}
\ee
The axion $a(x)$ relaxes towards the minimum of its potential, at  $ \langle a \rangle =0$, thus  explaining why $\bar{\Theta}$ is very small today.
 However, in the early universe,  just after $U(1)_{PQ}$ breaking, $\bar{\Theta}=a(x)/f_a$ is large and frozen to a value of order 1 as long as the axion is massless. 
 
 The axion acquires a mass at the QCD phase transition and classical oscillations of the axion background field around the minimum of the potential  start at $T_i$ when its mass is of the order of the Hubble scale, $m_a(T_i)\sim 3 H(T_i)\sim \Lambda_{QCD}^2/M_{Planck}$.
The energy stored in these axion oscillations redshifts as non relativistic matter, and eventually behave exactly as cold dark matter, according to
$
{d^2\langle a \rangle}/{dt^2} + 3 H(t) {d\langle a \rangle}/{dt} + m_a^2(t) \langle a\rangle =0
$
that leads to
$
\rho_a=\rho(T_i) \left[{m_a}/{m_a(T_i)}\right]\left[  {R^3(T_i)}/{R^3} \right]\sim \Lambda_{QCD}^3T^3/(m_a M_{Planck})$
 where $\rho(T_i)\sim m_a(T_i)^2f_a^2$. The energy density stored in axions today  is bounded by $\rho_{\tiny CDM}\sim 10^{-47} \mbox{GeV}^4 \gtrsim\bar{\Theta}^2m_a^2f_a^2/2 \sim \bar{\Theta}^2 m_{\pi}^2 f_{\pi}^2$. Therefore, today:
 $ \bar{\Theta} \lesssim 10^{-21}$.
While the PQ mechanism works independently of the value $f_a$ at which the symmetry is broken, the above cosmological constraint leads to an upper bound on $f_a$ whereas  an astrophysical lower bound  from excess cooling of supernovae, narrows  the remaining allowed window for the QCD axion, still unaccessible to laboratory experiments, to \cite{Dias:2014osa}:
\be
10^9 \mbox{ GeV} \lesssim f_a  \lesssim 10^{11} \mbox{ GeV}
\ee

The question we want to address in this paper is whether
 $\bar{\Theta}$ could have played any role at the time of the EW phase transition (EWPT).
Given that
the physical effects of $\bar{\Theta}$ are testified by the absence in strong interactions of  the isosinglet axial symmetry and its associated Goldstone,  a light pseudo scalar meson with mass comparable to pions
 \cite{'tHooft:1986nc}, and that  PQ solution is essentially the only solution to the strong CP problem, 
there are strong motivations for considering the  role of $\bar{\Theta}$ in the early universe. 
This question has been investigated only once in the literature in Ref.~\cite{Kuzmin:1992up}, by Kuzmin, Tkachev and Shaposhnikov who concluded that strong CP violation  could not be responsible for baryogenesis. 
 We revisit this question here and draw different conclusions from Ref.~\cite{Kuzmin:1992up}.
We show that this almost-SM source of CP violation can explain  baryogenesis under rather minimal assumptions.

A baryogenesis theory requires a stage of non-equilibrium dynamics in addition to CP-violation and baryon  number (B) violation.
One of the most popular route for baryogenesis has relied on the possibility that the EWPT is first-order. In this case, baryogenesis takes place in the wall vicinity of growing bubbles of EW-broken vacuum where $\langle \phi \rangle \neq 0$. In front of the bubble wall, in the symmetric phase, where $\langle \phi \rangle = 0$, sphalerons are active and B production is biased by CP-violating reflections of particles on the bubble wall. While the universe is converted into the EW broken phase where sphalerons are frozen, the B asymmetry cannot be washed out. This can only work if the EWPT is sufficiently first-order, i.e. $\phi/T \gtrsim 1$ at the time of bubble  nucleation. In the SM, the  EWPT is a crossover  and the system stays close to equilibrium but in models with an extended scalar sector, not only the EWPT can easily be first-order but new CP-violating sources also come into play. EW baryogenesis relies on new electroweak/TeV scale physics only and  is  testable at the LHC.

Another less-known but  interesting route is to consider instead 
 the case where EW symmetry breaking is triggered  through a  fast tachyonic 
 instability \cite{GarciaBellido:1999sv,Krauss:1999ng,Copeland:2001qw,Cornwall:2001hq}. 
In this case,  the Higgs mass squared is not turning negative as a  consequence of the standard cooling of the universe  but because 
 of its coupling to another scalar field field which is rolling down its potential. As the Higgs mass is ``forced" to change rapidly in an almost empty universe by the  vacuum expectation value (VEV) of an extra field, this case is labelled as  ``Higgs quenching''.
 It has been shown that Higgs quenching leads to the production of unstable EW field configurations ($SU(2)$ textures) which when decaying lead to Chern--Simons number transitions. This dynamics can lead to very efficient production of B at zero temperature, and is at the origin of the {\it  cold} baryogenesis scenario \cite{Smit:2002sn,GarciaBellido:2003wd,Tranberg:2003gi,vanTent:2004rc,vanderMeulen:2005sp,Tranberg:2006dg,Enqvist:2010fd}. 
The cold baryogenesis  scenario  requires
\begin{itemize}
\item large Higgs quenching to produce Higgs winding number in the first place
\item  unsuppressed CP violation at the time of quenching to bias a net Êbaryon number 
\item a reheat temperature below the sphaleron freese-out temperature $T \sim 130$ GeV \cite{D'Onofrio:2014kta}  to avoid washout of $B$ by sphalerons.
\end{itemize}

Sphalerons play no role in this mechanism. Baryogenesis takes place at low temperature, due to the out-of-equilibrium production of  SM EW large field configurations. 
If the conditions above are satisfied, cold baryogenesis can successfully  account for the value of $\eta$.
It has been simulated on the lattice for different Higgs quenching parameters \cite{Tranberg:2006dg} and using as 
CP-violating source  the dimension-6 operator
 \be
  \frac{\Phi^{\dagger} \Phi}{M^2}   \mbox{Tr } F \tilde{F} 
 \label{eq:usualCPviolation}
 \ee
 where $\Phi$ is the Higgs doublet and  $F$ is the EW field strength. The contribution of (\ref{eq:usualCPviolation}) to the electron EDM  $d_e/e \approx m_e \sin^2\theta_W/(8\pi^2 M^2) \log[(M^2+m_H^2)/m_H^2]$ \cite{Lue:1996pr} is constrained by
the latest electron EDM limit \cite{Baron:2013eja}, leading to a bound on the new physics scale $M\gtrsim 65$ TeV. This still leaves some room open for cold baryogenesis which requires $M\lesssim 340$ TeV  \cite{Tranberg:2006dg}.

In this work we show that strong CP violation from the operator (\ref{eq:axion})
could source cold baryogenesis instead of the operator (\ref{eq:usualCPviolation}).
Besides, we are interested in providing natural justification for the Higgs quenching.
So far it is put by hand in simulations, implicitly postulating some new scalar at the weak scale responsible for Higgs quenching,  without worrying about the associated hierarchy problems.
A first step in this direction was provided in Ref.~\cite{Konstandin:2011ds} where it was argued that cold baryogenesis should follow naturally in models with nearly conformal dynamics. 
We show that a Higgs-dilaton coupling not only  naturally delays the EWPT to temperatures  $T\lesssim\Lambda_{QCD}$ but also induces sufficient Higgs quenching during the EWPT,  while keeping the reheat temperature below the sphaleron freese out temperature,  therefore naturally enabling cold baryogenesis.

The plan of this paper is as follows.
In Section \ref{sec:SMbaryogenesis} we review how a B asymmetry can arise from a time-dependent coupling to the 
$F\tilde{F}$ operator. In Section \ref{sec:axionCP} we show how the axion can source this time-dependent coupling and how a sufficiently large B asymmetry can be produced 
in the context of cold baryogenesis.
In Section \ref{sec:quenching}, we define the conditions for Higgs quenching. In Section \ref{sec:dilaton} we present the constraints on the dilaton potential. In Section \ref{sec:DM} we discuss cosmological implications of the delayed EWPT, in particular for QCD axion dark matter,  and we conclude in Section \ref{sec:conclusion}.

\section{Baryogenesis from SM baryon number violation}
\label{sec:SMbaryogenesis}

Baryon number violation in the SM follows from the EW anomaly
\be
\partial_{\mu} j^{\mu}_{B}=N_F \partial_{\mu} j^{\mu}_{CS}= N_F \frac{\alpha_W}{8 \pi} \mbox{Tr} F \tilde{F}
\ee
where $N_F$ is the number of families, $F$ is the EW field strength and $N_{CS}=\int d^3x j_{CS}^0$ is the Chern--Simons number.
Variations in the baryon number  are related to variations in the Chern--Simons number by 
$\Delta B= N_F \Delta N_{CS}$.
The master equation for baryogenesis is of the form
\be
\dot{n}_{CS}=-\frac{\Gamma}{T} \frac{\partial {\cal F}}{\partial N_{CS}}= \frac{\Gamma}{T} \mu_{CS}
\ee
where $\Gamma$ is the rate of Chern--Simons transitions and $\mu_{CS}$ is the chemical potential from CP-violating sources inducing a non-vanishing B number. We have omitted in the right-hand side  the washout term proportional to $-  \Gamma n_{CS}T^{-2}$, assumed to be negligible.
The generated Chern--Simons number asymmetry can then be written
\be
\langle  N_{CS} \rangle (t) = \frac{1}{T_{eff}} \int_0^t dt' \  \Gamma_{CS}(t')  \  \mu(t')
\ee
where $T_{eff}$ characterizes the temperature at which Chern--Simons transitions are operative
at the same time as efficient CP-violation effects.

Relevant for baryogenesis is the effective lagrangian
\be
{\cal L}_{eff}=  \frac{\alpha_W}{8\pi}  \ \zeta(\varphi)  \mbox{Tr } F \tilde{F} 
\label{eq:keyoperator}
\ee
where $\zeta(\varphi)$ is some time-varying function of fields which depends on the underlying baryogenesis model. 
We have
\be
\int  d^4x \frac{\alpha_W}{8\pi}  \ \zeta  \ \mbox{Tr } F \tilde{F} =\int  d^4x  \ \zeta \ \partial_{\mu} j^{\mu}_{CS}=-
\int  dt \ \partial_t {\zeta} \int d^3xj^0_{CS}
\ee
where we made an approximation in which $\zeta$  is replaced by its spatial average $L^{-3}\int d^3x \zeta$ and we integrated by parts in order to exhibit the chemical potential for Chern--Simons
number:
\be
\mu \equiv \frac{d}{dt} \zeta(t)
\ee
Therefore, the time derivative of $\zeta$ can be interpreted as a time-dependent chemical potential for Chern--Simons number and ${\cal L}_{eff}$ takes the form
 \be
 {\cal L}_{eff}= \mu \ N_{CS}.
 \ee
This has been extensively used in baryogenesis scenarios in the past.
The produced B asymmetry is given by
\be
n_B =N_F \int dt \frac{\Gamma  \mu} {T} \sim N_F \frac{\Gamma(T_{eff})}{T_{eff}} \Delta \zeta
\ee
Using the sphaleron rate in the EW symmetric phase 
\be
N_F\Gamma=30 \alpha_w^5 T^4\sim \alpha_w^4 T^4,
\label{eq:sphaleronrate}
\ee
 this leads to:
\be
\frac{n_B}{s}=N_F  \alpha_w^4 \left(\frac{T_{eff}}{T_{reh}}\right)^3 \ \Delta \zeta  \ \frac{45}{2 \pi^2 g_*(T_{reh}) } 
\sim 10^{-7} \left( \frac{T_{eff}}{T_{reh}}\right)^3 \Delta \zeta
\label{eq:baryonasymmetry}
\ee
where $T_{reh}$ is the reheat temperature after the EWPT and is of the order of the Higgs mass. It may be significantly higher than the temperature of the EWPT, $T_{EWPT}$, if the EWPT was delayed and completed after a supercooling stage \cite{GarciaBellido:1999sv}.
For standard EW baryogenesis, $T_{eff}=T_{EWPT}=T_{reh}$.
 In contrast, the key-point for cold baryogenesis is that $ T_{eff} \neq T_{EWPT} $ \cite{GarciaBellido:1999sv}. 
 $T_{eff}$ should be viewed as an effective temperature associated with the production of  low-momentum Higgs modes during quenching. It is significantly higher than the temperature of the EWPT. It is a way to express the very efficient rate of B violation in terms of the equilibrium expression $ \Gamma \sim \alpha_w^4 T^4$ although the system is very much out-of equilibrium. We will come back to this in Section \ref{sec:quenching}.
 
 As a new source of CP violation, most studies of cold baryogenesis have relied on the effective dimension-6 operator given in Equation (\ref{eq:usualCPviolation}), i.e.
 \be \zeta=\frac{8\pi}{\alpha_W} \frac{\Phi^{\dagger} \Phi}{M^2}
 \label{eq:usualzeta}
 \ee 
The resulting B asymmetry is given by
\be
\frac{n_B}{s} \sim 10^{-5} \left(\frac{T_{eff}}{T_{reh}}\right)^3 \frac{v^2}{M^2}
\ee
Using $T_{eff} \gtrsim 5 T_{reg}$ \cite{Tranberg:2006dg}, we can have a large enough asymmetry in the context of cold baryogenesis and satisfy the bound $M\gtrsim 65$ TeV from the  electron electric dipole moment.

In summary, the time-varying VEV of the Higgs field enables successful cold baryogenesis.
What we are instead going to use in our proposal is that $\zeta$ is actually fueled by the time variation of the axion mass around the QCD scale, while the rate of Chern--Simons transitions is non-zero because of the EWPT being delayed at or below the QCD scale in the context of dilaton-induced EW symmetry breaking.

\section{Axion-induced CP violation}
\label{sec:axionCP}

Our goal is to investigate whether the large values of the effective vacuum angle  in Eq.~\ref{eq:axion}
 at early times can have any implications for EW baryogenesis.
We have 
\be
\bar{\Theta}={a}/{f_a} \sim {\cal O}(1) \mbox{ for } T\gtrsim 1 \mbox{ GeV},
\ee
and then $\bar{\Theta}$ quickly drops as the axion gets a mass and starts oscillating around the minimum of its potential. 
The axion lagrangian reads:
 \be
 {\cal L}_a= {\cal L}(\partial_{\mu}a) -\frac{1}{2}\partial^{\mu}a\partial_{\mu}a +\frac{a}{f_a}  \frac{\alpha_s}{8\pi} G\tilde{G}
 \ee
 so that
 \be
 \frac{\partial V_{eff}} {\partial a}=-\frac{1}{f_a} \frac{\alpha_s}{8\pi} G\tilde{G}
 \ee
 Gluon condensation from $SU(3)$ instantons leads to a VEV for $G\tilde{G}$ and a potential for the axion that can be written as
 \be
 V=f^2_{\pi}m^2_{\pi} (1-\cos \frac{a}{f_a})\approx f^2_{a}m_a^2 (1-\cos \frac{a}{f_a}).
 \ee
As a result
\be
 \frac{\alpha_s}{8\pi} \langle G\tilde{G}  \rangle=f_a^2 m_a^2 \sin \bar{\Theta}.
\ee
To make a connection between the axion and EW baryogenesis, we have to construct
 an effective operator gathering gluons and EW gauge bosons. 
The main point of the previous section can be summarized as 
\be
{\cal L}_{eff}= \frac{\alpha_W}{8\pi}  \ \zeta(T)  \mbox{Tr } F \tilde{F}  \ \  \leftrightarrow  \ \  {\cal L}_{eff}= \mu N_{CS} \ \ \mbox{where} \ \ \mu=\frac{d}{dt} \zeta(T) 
\ee
An operator of the type (\ref{eq:keyoperator}) can arise, where $\zeta$ is controlled by the axion mass squared.
 In particular, the $\eta^{\prime} $ meson, which is a singlet under the approximate $SU(3)$ flavor symmetry of strong interactions, can couple to both $G\tilde{G}$ and $F\tilde{F}$. At temperatures below the $\eta^{\prime}$ mass, $m_{\eta^{\prime}}\approx 958$ MeV, we can use the effective operator
\be
{\cal L}_{eff}=
\frac{1 }{M^4} 
  \  \frac{ \alpha_s }{8 \pi}  G\tilde{G} \  \ \ \frac{\alpha_w}{8\pi} F\tilde{F}
\ee
where $1/M^4=10/({F_{\pi}^2 m_{\eta^{\prime}}^2} )$ \cite{Kuzmin:1992up}. 
%
We end up with 
\be
{\cal L}_{eff}=\frac{1 }{M^4} \  \sin \bar{\Theta} \ m_a^2(T) f_a^2  \ \ \frac{\alpha_w}{8\pi} F \tilde{F}
\ee
hence
\be
\zeta(T)\equiv \frac{1 }{M^4} \  \sin \bar{\Theta} \ m_a^2(T) f_a^2  \ \rightarrow \ \mu=\frac{d \zeta}{dt}=\frac{f_a^2 }{M^4} \ \frac{d }{dt}[ \sin \bar{\Theta}  \ m_a^2(T) ]
\ee
As announced earlier, the time variation of the axion field and/or mass is a source for baryogenesis:
\be
n_B  \propto  \int dt \frac{\Gamma(T)  } {T} \frac{d}{dt} [\sin \bar{\Theta} \ m^2_a(T)] 
\ee
To estimate the resulting B asymmetry, we will use as the temperature-dependent axion mass
with $T_t=102.892$ MeV  \cite{Wantz:2009it}:
\be
f^2_a m_a^2(T)= \frac{\alpha_a \Lambda^4}{\left({T}/{\Lambda}\right)^{6.68} } \  ,  \mbox{ for } T>T_t ,\mbox{ where } \Lambda=400 {\mbox{ MeV and }} \alpha_a=1.68 \times 10^{-7}
\ee
which we can rewrite as
\bea
\nonumber
m_a^2(T) &=& m_a^2(T=0)  \  ,  \mbox{ for } T\leq T_t  \\
m_a^2(T) & \approx & m_a^2(T=0) \times \left(\frac{T_t}{T}\right)^{6.68} \  ,  \mbox{ for } T > T_t 
\label{eq:Tmass}
\eea
where $m_a^2(T=0)f_a^2=m_{\pi}^2 f_{\pi}^2  {m_u m_d}/{(m_u+m_d)^2}$.
The axion mass is very suppressed at temperatures above the QCD scale. A large B asymmetry is therefore produced only if the EWPT  occurs not much earlier than the QCD phase transition.
We have
\be
\Delta \zeta \sim  \left. \frac{10 f_a^2}{f_{\pi}^2 m_{\eta^{\prime}}^2} \delta  [\sin \bar{\Theta}  \ m^2_a]\right|_{EWPT} 
 \sim 0.044  \sin \bar{\Theta}(T_{EWPT})\times  \left(\frac{T_t}{T_{EWPT}}\right)^{6.68} \  ,  \mbox{ for } T_{EWPT} > T_t  
\ee
where we used $\delta \sin \bar{\Theta} \ m_a^2(T)\sim \sin \bar{\Theta} \ m_a^2(T) $. So the final B asymmetry in the absence of washout, is, using Eq (\ref{eq:baryonasymmetry}):
\bea
\nonumber
\frac{n_B}{s}&\sim & 4 \times 10^{-9}  \left(\frac{T_{eff}}{T_{reh}} \right)^3 \sin \bar{\Theta}(T_{EWPT})  \  ,  \mbox{ for } T_{EWPT} < T_t  \\
\frac{n_B}{s}&\sim &4 \times 10^{-9}  \left(\frac{T_{eff}}{T_{reh}} \right)^3 \sin \bar{\Theta}(T_{EWPT}) \left(\frac{T_t}{T_{EWPT}}\right)^{6.68}  \  ,  \mbox{ for } T_{EWPT} > T_t  
\label{eq:axionbau}
\eea
If we work in the context of standard EW baryogenesis, we have  $T_{eff}=T_{EWPT}=T_{reh}$.
Even if we set $ \sin \bar{\Theta}(T_{EWPT}) \sim 1$ in Eq.~\ref{eq:axionbau},  we need $ T_{EWPT}\lesssim 0.2 $ GeV, to get a large enough B asymmetry today.  Such a low EWPT  temperature can be achieved by coupling the Higgs to a dilaton field \cite{Buchmuller:1990ds,Konstandin:2011dr,Konstandin:2011ds}  whose scalar potential energy induces a supercooling stage\footnote{Another option for a delayed EWPT considered by Witten in the eighties when it was still experimentally allowed was to  assume a Higgs mass $m_H\sim {\cal{O}}(1)$ GeV  \cite{Witten:1980ez}.}. 
However, the reheat temperature $T_{reh}\sim {\cal O} ({m_H,m_d})$  cannot be kept below a GeV, unless the dilaton with mass $m_d$ is very weakly coupled.
Besides, in this case, axion oscillations would be delayed and would overclose the universe. This led Kuzmin et al. to conclude that strong CP violation from the axion cannot play any role during EW baryogenesis \cite{Kuzmin:1992up}. We are going to conjecture on the contrary that the axion can well explain baryogenesis.

\begin{figure}[t!]
\begin{center}
\includegraphics[angle=0,width=0.6\linewidth]{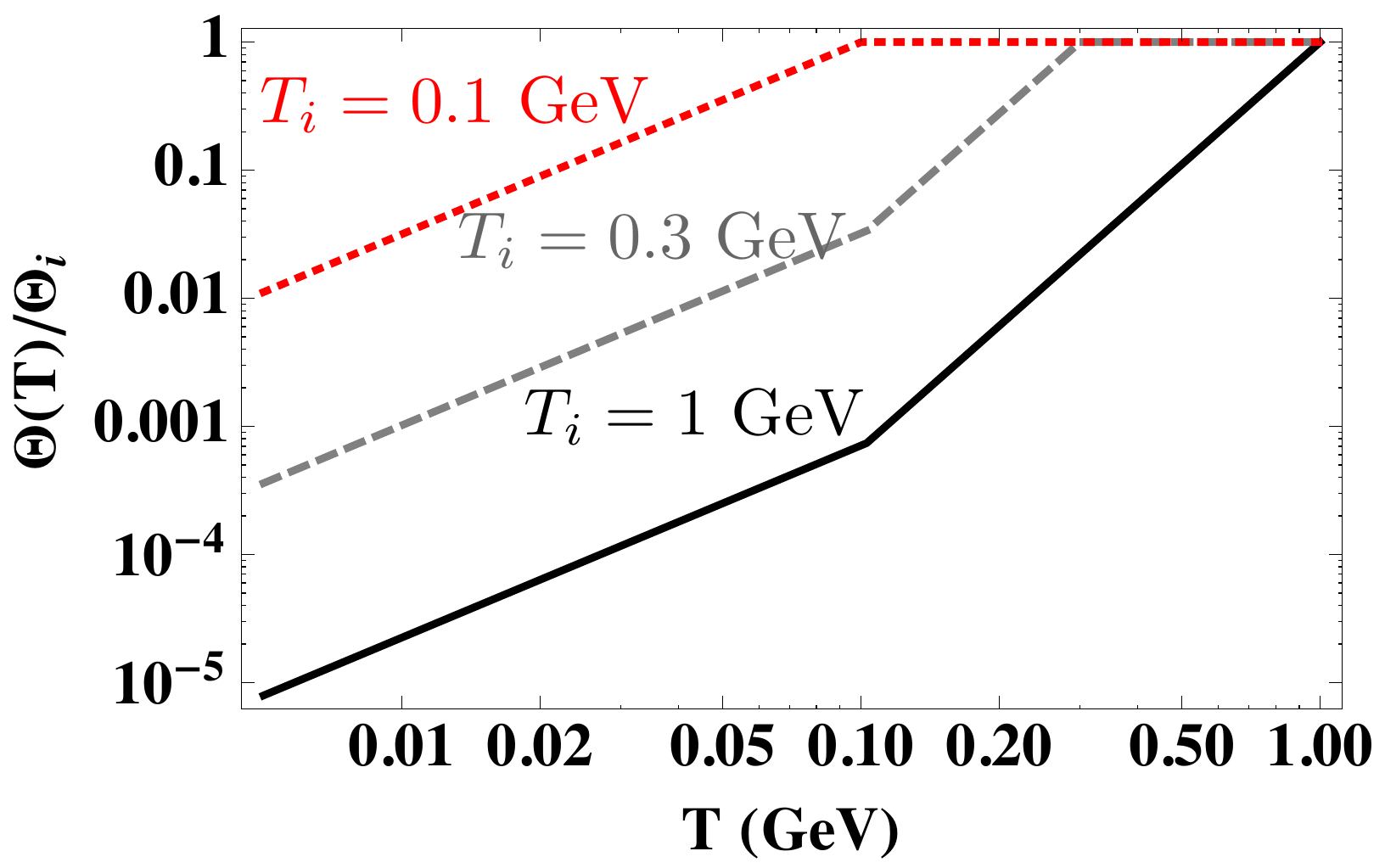}
\caption{\small Evolution of $\bar{\Theta}/\bar{\Theta}_i$ assuming axion oscillations start respectively at  $T_i =0.1, 0.3, 1$ GeV, $\bar{\Theta}_i$ being the initial value. From Eq.(~\ref{eq:axionbau}),  the B asymmetry is proportional to $\bar{\Theta}(T_{EWPT})$, where $T_{EWPT}$ is  the  temperature of the EW phase transition. }
\label{fig:thetaevolution} 
\end{center}
\end{figure}

The B asymmetry given by  Eq.~\ref{eq:axionbau} is proportional to $\bar{\Theta}(T_{EWPT})$, which can be smaller than the initial value $\bar{\Theta}_i$ if axion oscillations started before the EWPT. Therefore, the B asymmetry  depends on the axion mass $m_a \propto 1/f_a$ which determines whether oscillations start before or after the EWPT. 
The temperature $T_i$ at which oscillations start  is defined
 by $3 H=m_a(T_i)$. In a radiation-dominated universe, $T_i\sim$ a few GeV. In the case of supercooling at the EW scale, $T_i$ is smaller, as will be discussed in Section \ref{sec:DM}.
 The energy density stored in axion oscillations redshifts as non-relativistic matter, leading to:
\be
\bar{\Theta}^2(T)= \bar{\Theta}^2_i  \  \frac{m_a(T_i)}{m_a(T)} \ \left(\frac{T}{T_i}\right)^3 \ \ \mbox {for} \ \ T\leq T_i
\label{eq:thetaofT}
\ee
The corresponding evolution of $\bar{\Theta}$  is plotted in Fig.\ref{fig:thetaevolution} assuming three different values for $T_i$. 
%
%

\begin{figure}[t!]
\begin{center}
\includegraphics[angle=0,width=0.999\linewidth]{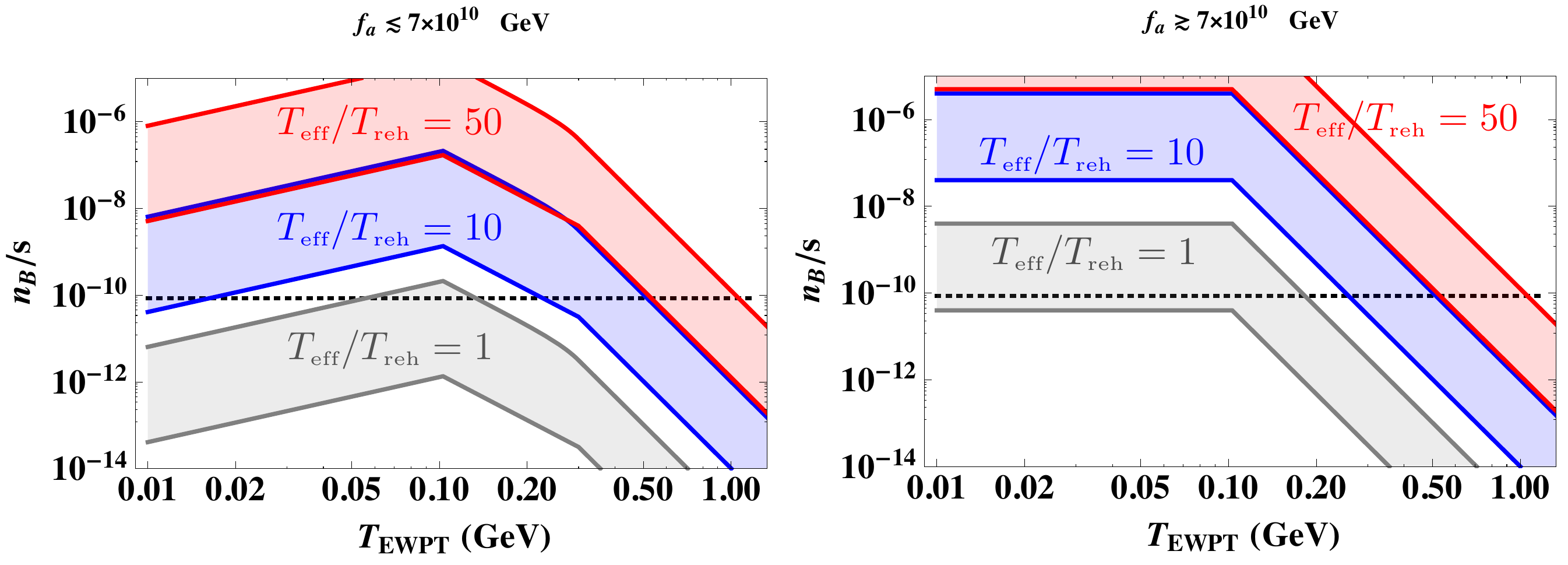}
\caption{\small 
 Prediction for  today's B asymmetry as a function of the temperature of the EWPT compared with measured value (dotted line).
The case $T_{\mbox{\tiny eff}}/T_{\mbox{\tiny reh}}=1$ (light gray) and $T_{\mbox{\tiny eff}} \sim T_{\mbox{\tiny EWPT}}$  that would characterize {standard} EW baryogenesis is  unfeasible  as $T_{reh}\sim {\cal O}(m_H) \gg \Lambda_{QCD}$. The cases with  $T_{\mbox{\tiny eff}}/T_{\mbox{\tiny reh}} \gtrsim 10$  can easily account for a large B asymmetry and correspond to a { quenched} EWPT, as in {cold} EW baryogenesis. Each band corresponds to varying the initial angle value $\bar{\Theta}_i$ in the range $[10^{-2},\pi/2]$. Left: $m_a\gtrsim 3H_{EW}\sim3\times 10^{14}$ GeV, for oscillations starting at $T=0.3$ GeV in the supercooling era before the EWPT. Right:  $m_a\lesssim 3H_{EW}$, the axion is frozen to its initial value  until after reheating.}
\label{fig:bau} 
\end{center}
\end{figure}

To deduce the B asymmetry we then plug (\ref{eq:thetaofT}) into (\ref{eq:axionbau}). The result is shown in Fig.~\ref{fig:bau}, in the absence of washout. For temperatures above the QCD phase transition, there is a suppression from the axion mass, while at low temperatures, the suppression comes from the smallness of  $\bar{\Theta}$ if the axion started to oscillate in the supercooling stage, i.e. $m_a\gtrsim 3H_{EW}\sim \sqrt{3\rho_{vac}}/ m_{Pl}$. In that case, the B asymmetry is then typically maximized for EWPT temperatures in the 10 MeV$-$1 GeV range.
In the left-handed plot, the gray curve, which corresponds to the standard EW baryogenesis assumption,  $T_{eff}=T_{reh}=T_{EWPT}$, reproduces  the negative conclusion by Kuzmin et al.
At the time of Ref.~\cite{Kuzmin:1992up}, the mechanism of cold baryogenesis was not known.  On the other hand, cold baryogenesis cures the problem. The key point is that even if $T_{EWPT} \lesssim \Lambda_{QCD}$, we can have 
$T_{eff} \gtrsim T_{reh}\sim m_H$.  From lattice  simulations of cold baryogenesis \cite{Tranberg:2006dg},
a { quenched} EWPT typically has 
\be
\frac{T_{eff}}{T_{reh}} \sim 20-30
\ee
Considering that there is no washout factor, we easily get a large B asymmetry since
${n_B}/{s}\propto ({T_{eff}}/{T_{reh}})^3$. For $m_a\lesssim 3H_{EW}$, there is no  low-temperature suppression as the axion field value does not start decreasing
until after the EWPT so that a large B asymmetry can be produced even for a very delayed EWPT, see right-handed plot in Figure.~\ref{fig:bau}.
The final result depends on the initial angle $\bar{\Theta}_i$ before axion oscillations start. We indicated  this dependence by a band corresponding to the range
 $ 10^{-2} \leq \bar{\Theta}_i \leq \pi/2$. 
We conclude that the standard QCD axion can be responsible of the B asymmetry of the universe in the context of cold EW baryogenesis.
We now review the conditions for successful cold EW baryogenesis.

\section{The Higgs quench from a Higgs-scalar coupling}
\label{sec:quenching}

The key point in this work is to exploit the fact that efficient B violation can take place at temperatures below the sphaleron freese-out temperature, under strong out-of-equilibrium conditions as provided by a quenched EWPT.
We summarize here briefly the main features of cold baryogenesis and refer the reader to the specific literature for more details \cite{GarciaBellido:1999sv,Smit:2002sn,GarciaBellido:2003wd,Tranberg:2003gi,vanTent:2004rc,vanderMeulen:2005sp,Tranberg:2006dg,Enqvist:2010fd}. 

 In the standard picture of cold baryogenesis, the tachyonic transition develops when the Higgs mass squared $m_{eff}^2$ changes sign rapidly due to a coupling of the Higgs to an additional scalar field. Just before the EWPT, the universe is relatively cold.
The dynamics of spinodal decomposition has been investigated both analytically and numerically 
\cite{GarciaBellido:2003wd,vanderMeulen:2005sp,Asaka:2001ez,Copeland:2002ku,GarciaBellido:2002aj,Smit:2002yg,Borsanyi:2003ib}, typically using infinitely fast quench. 
The Fourier modes of the Higgs field with low momentum $k<\mu$ are unstable and grow exponentially. The rapid rise of the low momentum modes  and  the particle number distribution of the Higgs can be seen by solving
$\ddot{\Phi}(k,t)+(m_{eff}^2(t)+k^2)\Phi(k,t)=0$ and assuming instantaneous quenching:  $m_{eff}^2=+\mu^2$ at  $t<0$ and $m_{eff}^2=-\mu^2$ at $t>0$, $t=0$ being the onset of the transition. This leads to $\Phi(k,t)\propto \exp[\sqrt{\mu^2-k^2} t]$. 
Therefore, the energy of the additional scalar field inducing the quench  is converted into long wavelength modes of the Higgs field which then contain a large fraction of the total energy of the system.
These extended field  configurations play a key role in inducing Chern--Simons transitions (see e.g. \cite{Konstandin:2011ds} for a summarized review and references therein).
It is difficult to predict the final averaged Chern--Simons number analytically.
On the other hand, although we are far from thermal equilibrium, we  can use some effective sphaleron rate to roughly estimate the effect of dilaton-induced baryon-number violation.
The rate of Chern--Simons transitions  can be approximated by  that of  a system in thermal equilibrium at a temperature $T_{eff}$.   
The effective temperature can be estimated comparing  Eq.~(\ref{eq:sphaleronrate})
with 
\be
\Gamma(t')=\frac{d   (\langle N^2_{CS}   (t)   \rangle  }{dt}
\ee
where $N_{CS}(t)=\int_0^t dt \int d^3x (16 \pi)^{-1} TrF\tilde{F}$.
The production of non-zero Chern-Simons number when the Higgs field experiences a fast quench 
  depends on the Higgs mass  and on the speed of the quench. 
 For $m_H \sim 125$ GeV and  quenching parameter $|u| \gtrsim 0.1$, lattice simulations
   \cite{Tranberg:2006dg} found
${T_{eff}}/{T_{reh}} \sim 20 -30$.

The quenching time $t_q$ is defined as the time when the Higgs mass turns negative.
The speed of the quench  or quenching parameter 
is a dimensionless velocity parameter characterizing  the rate of change of the effective Higgs mass squared at the time of quenching:
\be
u \equiv \left.  \frac{1}{ 2 m_H^3}\frac{d m_{\mbox{\tiny eff}}^2}{dt}\right|_{T=T_q} 
\ee
Successful cold baryogenesis requires $|u| \gtrsim 0.1$ \cite{Tranberg:2006dg}.
In the SM, the effective Higgs mass varies solely because of the cooling of the universe. 
Using $d/dt = - H T d/dT$ and  $T_q \sim \mu \sim m_H$,  the quenching parameter is then 
\be
u^{\mbox{\tiny SM}} \sim
\left. \frac{1}{\mu^3}\frac{d}{dt} ( -\mu^2  + c T^2) \right|_{T=T_q}
\sim \left. \frac{H}{\mu} \right|_{T_q} 
\sim \frac{T_{\mbox{\tiny EW}}}{M_{Pl}} \sim 10^{-16} 
\ee
This situation can be changed radically if the Higgs mass is controlled by the time-varying VEV of an additional field $\sigma$, {\it e.g.} 
\be
m_{\mbox{\tiny eff}}^2(t)=\mu^2-\lambda_{\sigma \phi} \sigma^2(t).
\label{eq:standard mueff}
\ee
 Then 
\be
u=- \frac{{\lambda_{\sigma \phi}}}{m_H^3}
 \left. (\sigma \dot{\sigma}) \right|_{t_q}.
 \label{eq:quenching}
\ee
 From energy conservation $(\dot{\sigma})^2 \sim {\cal O}(\Delta V)$ and we can naturally get order 1 quenching parameter as it is no longer controlled but the Hubble parameter. 
This additional coupling of the Higgs is what the cold baryogenesis scenario assumes. In this paper, we provide a natural motivation for such an assumption. Earlier proposals rely on arbitrary potentials  in which the masses of the scalars are not protected. Instead, we show that the mechanism can be implemented in a well-motivated framework where the smallness of the scalar masses is under control.
We already made these claims in Ref.~\cite{Konstandin:2011ds}.
In the following we build up on this work.

\section{Naturally delayed EWPT and low reheat temperature from the dilaton}
\label{sec:dilaton}

We consider the following scalar potential in which the quadratic term for the Higgs field $\phi$ is controlled by the 
VEV of the dilaton $\sigma$:
\be
V(\sigma,\phi)= V_{\sigma}(\sigma) +\frac{\lambda}{4} (\phi^2-\xi \ \sigma^2)^2
\label{eq:higgsdilatonpotential}
\ee
where $V_{\sigma}(\sigma) $ is  a scale invariant function  modulated by a slow evolution:
\be
V_{\sigma}(\sigma)= \sigma^4 \times P(\sigma^{\epsilon}) \ \ \mbox{where } \ |\epsilon| \ll 1
\label{eq:dilatonpotential}
\ee 
 and $\xi$ is a constant.
Note that this potential is precisely the one of Randall-Sundrum models \cite{Randall:1999ee}. Assuming the Higgs is localized on the IR brane at a distance $y=r$ from the UV brane localized at $y=0$,  the 4D  effective action for the Higgs is:
\be
{\cal L}_4=e^{-2k\pi r} \eta^{\mu\nu} D_{\mu}\tilde{H} D_{\nu} \tilde{H} -e^{-4k\pi r} \lambda(|\tilde{H}|^2-v_P^2)^2
=\eta^{\mu\nu} D_{\mu}{H} D_{\nu} {H} - \lambda \left(|{H}|^2-(\frac{v_P}{k} \sigma)^2 \right)^2
\ee
where $v_P \sim \Lambda_{UV}\sim m_{Pl}\sim k$, $H$ is the canonically normalized field $H=e^{-k \pi r} \tilde{H}$ and  the radion field is
\be
\sigma \equiv k e^{-k \pi r}.
\ee
We also define the scale
\be
 f \equiv \langle \sigma \rangle 
\ee
which is generated once the radion is stabilized and is exponentially warped down from the Planck scale due to the Anti de Sitter geometry. We also have
$\xi=v^2/f^2$.
For the 5D AdS metric, the effective 4D potential for the radion was shown  to be of the dilaton-like form (\ref{eq:dilatonpotential}), 
independently of the inter brane distance stabilization mechanism \cite{Goldberger:1999uk,Rattazzi:2000hs,Garriga:2002vf}. We therefore recover the scalar potential (\ref{eq:higgsdilatonpotential}) for the coupled radion-Higgs system. Solving the weak/Planck scale hierarchy leads to $f \sim {\cal O}$(TeV).

The cosmology of the potential (\ref{eq:higgsdilatonpotential}), $V_{\sigma}(\sigma)$, was summarized in Ref.~\cite{Konstandin:2011dr}.
A very strong first-order phase transition typically occurs for this type of potential. 
In the first investigation of the associated phase transition  it was argued that transition to the minimum of the radion potential could not complete \cite{Creminelli:2001th}.  This conclusion essentially followed from a thin-wall estimate of the critical bubble action and assuming that tunneling would take place directly to the minimum of the potential.
The key point stressed in \cite{Randall:2006py} is that the phase transition can actually complete through tunneling to a value of the field much smaller than the value at the minimum of the potential and the field subsequently rolls towards the minimum.
This is typical of very shallow potentials.
As emphasized in Ref.~\cite{Konstandin:2011dr}
the value of the field at tunneling, $\sigma_r$,  is
\be 
\sigma_r \sim  \sqrt{ \sigma_{+ }\sigma_{-}}
\ee
where $\sigma_{+}$ and  $\sigma_{-}=f$ are the positions of the maximum and minimum of the potential respectively.
The nucleation temperature $T_n$ is proportional to $\sigma_r$ and given by \cite{Konstandin:2011dr,Konstandin:2010cd}
\be
T_n \sim 0.1 \sqrt{\sigma_{+ }\sigma_{-}} \sim 0.1 \ f \  \sqrt{\frac{\sigma_+ }{\sigma_{-}}}
\ee
For a standard polynomial potential, $\sigma_+ \sim \sigma_- \sim \sigma_r \sim T_n$. In contrast, for the very shallow dilaton-like potential, $\sigma_+ \ll \sigma_-$, and the nucleation temperature is parametrically much smaller than the scale associated with the minimum of the potential. We therefore naturally get a   stage of supercooling before the phase transition completes. The hierarchy between $\sigma_{-}$ and $\sigma_{+}$ can be as large as the Planck scale/weak scale hierarchy:
\be
\frac{\sigma_{-}}{\sigma_{+}} \lesssim \frac{\Lambda_{UV}}{\Lambda_{IR}}
\ee
 Therefore the nucleation temperature can  be as low as ~\cite{Konstandin:2011dr}
\be
T_n\sim 0.1  \Lambda_{IR}\sqrt{\Lambda_{IR}/\Lambda_{UV}}
\ee
We obtain  $T_n\sim 35$  MeV if 
$ \Lambda_{IR}=5 $ TeV and  $\Lambda_{UV}=M_{Pl}$, while
  $T_n\sim 0.1$  GeV if 
$ \Lambda_{IR}=1 $ TeV and  $\Lambda_{UV}=f_{a}=10^{10}$ GeV.
Note that as QCD breaks conformal invariance, it modifies the potential (\ref{eq:dilatonpotential}). While a delayed EWPT down to the QCD scale  is a general outcome in our framework,  the modification of the scalar potential around the QCD scale will affect the detailed predictions.
%
\begin{figure}[t!]
\begin{center}
\includegraphics[angle=0,width=0.6\linewidth]{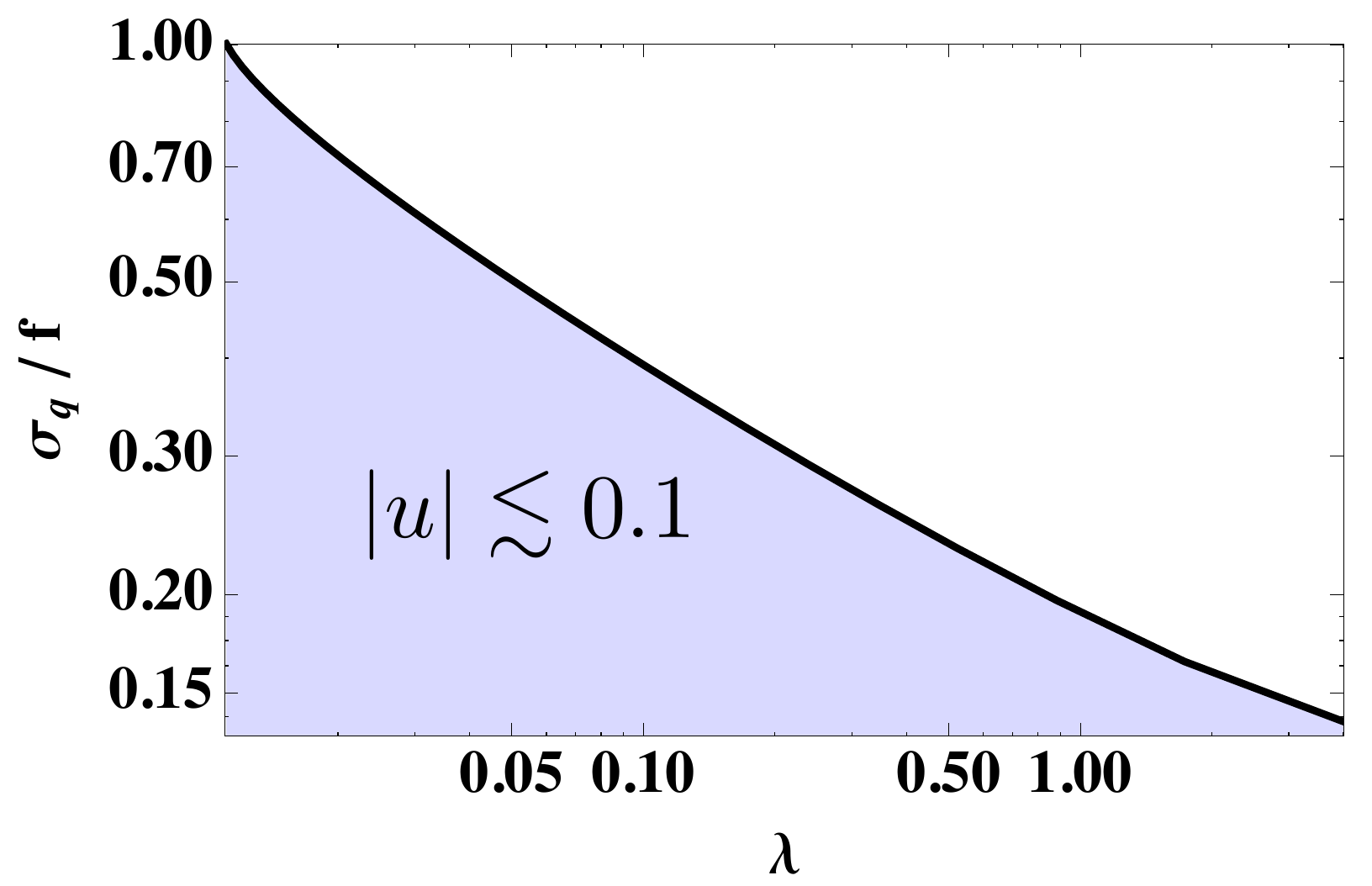}
\caption{\small Value of the dilaton field for which the cold baryogenesis quenching condition is satisfied as a function of the Higgs quartic coupling $\lambda$, for a typical dilaton potential (we used the Goldberger-Wise potential from  \cite{Randall:2006py} for this illustration). As $\sigma$ approaches the minimum $\langle \sigma \rangle =f$, the quenching condition is satisfied even for relatively small values of $\lambda$. }
\label{fig:quenching} 
\end{center}
\end{figure}

In summary, the temperature of the phase transition associated with a potential (\ref{eq:dilatonpotential}) can be many orders of magnitude smaller than the scale given by the VEV at the minimum of the potential.
As $v\ll f$, the Higgs dependence in the scalar potential was ignored in  Ref.~\cite{Konstandin:2011dr}, which focussed on $V_{\sigma}(\sigma)$.
It was implicitly assumed that 
out-of-equilibrium Higgs dynamics would follow automatically from the radion/dilaton dynamics, in particular  as a consequence of bubble collisions, generating the conditions to produce Higgs winding configurations, as needed for cold baryogenesis. As a proof of principle we simulated bubble collisions and looked at the transfer of potential energy to kinetic energy of the scalar field.

In the following, we examine more closely the quenching condition for the potential (\ref{eq:higgsdilatonpotential}). 
In the cases considered so far for cold baryogenesis, the quenching time is defined when the effective mass 
(\ref{eq:standard mueff}) vanishes, which translates as $\sigma^2_{q}=\mu^2/\lambda_{\sigma \phi}$. 
For the dilaton-like potential, $\mu=0$, and quenching should happen between tunneling and the time
the field rolls down to the minimum.
From (\ref{eq:quenching}), the condition $u\gtrsim 0.1$ becomes
\be
\frac{\xi \lambda}{m_H^3} \sqrt{2(V_{\sigma}(0)-V_{\sigma}(\sigma))} \ \sigma \gtrsim 0.1
\label{eq:quenchingminimum}
\ee
which we plot in Figure \ref{fig:quenching}. The condition on $\lambda$ is weakest at the minimum of the potential. Depending on the value of $\lambda$, the quenching condition can be realized relatively well before the field reaches the minimum. When estimated at the minimum, the condition (\ref{eq:quenchingminimum}) translates as a bound on the second scalar mass eigen state:
\be
 \frac{\lambda v^2 m_d}{m_H^3} \sim \frac{m_d}{2 m_H}
 \gtrsim 0.1
\ee
where $m_d=\sqrt{ V_{\sigma}^{''} (\sigma=f)}$.
Since we are considering $\xi=v^2/f^2 \ll 1$ the off-diagonal terms in the squared mass matrix of the Higgs-dilaton system are  small compared to the 
diagonal entries. Therefore the two mass eigen values are essentially $2\lambda v^2$ and $V_{\sigma}''(f)$.
%
%
\begin{figure}[t!]
\begin{center}
\includegraphics[angle=0,width=0.6\linewidth]{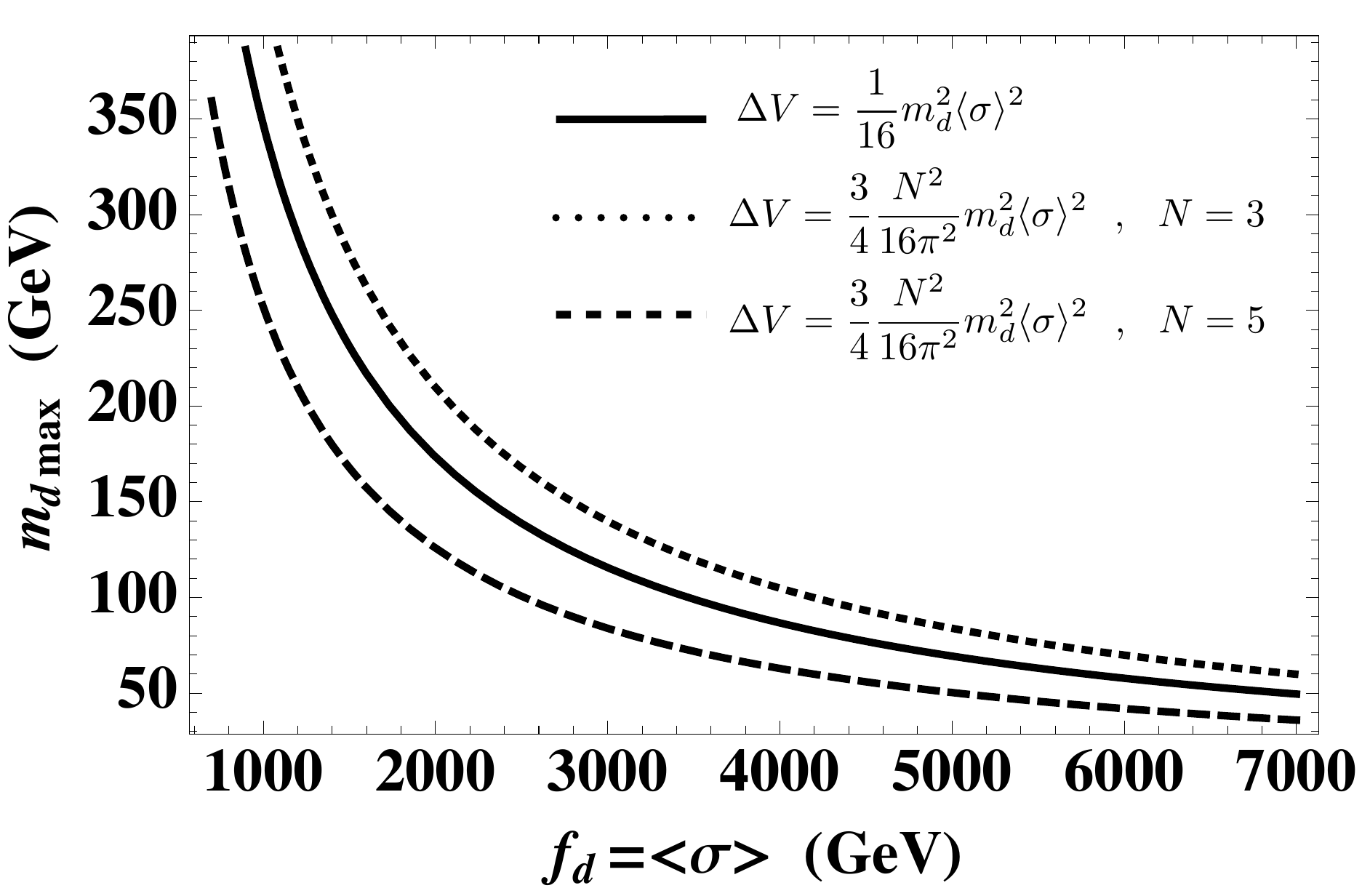}
\caption{\small Upper bound on the dilaton mass for the reheat temperature to be below the sphaleron freese-out temperature, $T=130$ GeV,  as a function of  the dilaton VEV $ \langle \sigma\rangle$, for three choices of dilaton potentials, taken respectively from Ref.~\cite{Bellazzini:2013fga} (solid line) and Ref.~\cite{Konstandin:2011dr} (dotted and dashed lines).}
\label{fig:Treheat} 
\end{center}
\end{figure}

Having checked that the quenching criterion is readily satisfied, the last condition to be satisfied
 for successful cold baryogenesis is that the reheat temperature is sufficiently small to prevent sphaleron washout.
After the EWPT, the vacuum energy stored in the Higgs and dilaton fields reheats the plasma.
\be
\frac{8 \pi g_* T^4_{reh}}{30} =\Delta V
\ee
where
\be
\Delta V \sim m_d^2 f^2
\ee
We have to impose that this temperature never exceeds the sphaleron freese-out temperature \cite{D'Onofrio:2014kta}
\be
T_{reh} < 130 \mbox{ GeV}
\label{eq:freeseoutTbound}
\ee
which leads to a constraint on the dilaton mass. Since $f\sim {\cal O}(TeV)$, this means that  the 
dilaton should be ${\cal O}(100)$ GeV. We plot this constraint in Fig.~\ref{fig:Treheat} for typical dilaton-like potentials used in the literature.
Constructions that lead naturally to such a  light dilaton have been recently discussed in 
Ref.~\cite{Rattazzi:2000hs,Coradeschi:2013gda,Bellazzini:2013fga,Megias:2014iwa,Chacko:2013dra}.
LHC constraints on an EW scale  dilaton  were presented before the Higgs discovery in 
\cite{Coleppa:2011zx,Barger:2011hu,Campbell:2011iw,deSandes:2011zs}.
Interpretation of the Higgs discovery in terms of a Higgs-like dilaton \cite{Goldberger:2008zz}  has then been considered in \cite{Bellazzini:2012vz,Chacko:2012vm}. We are  instead interested in a scenario where in addition to the 125 GeV Higgs, there is a light dilaton, which is a less constrained option,  see e.g \cite{Desai:2013pga,Cao:2013cfa,Jung:2014zga}, and a careful analysis of CMS and ATLAS data is generally definitely worthwhile and will be a key-test for our scenario in particular. 

%
\begin{figure}[t!]
\begin{center}
\includegraphics[angle=0,width=0.6\linewidth]{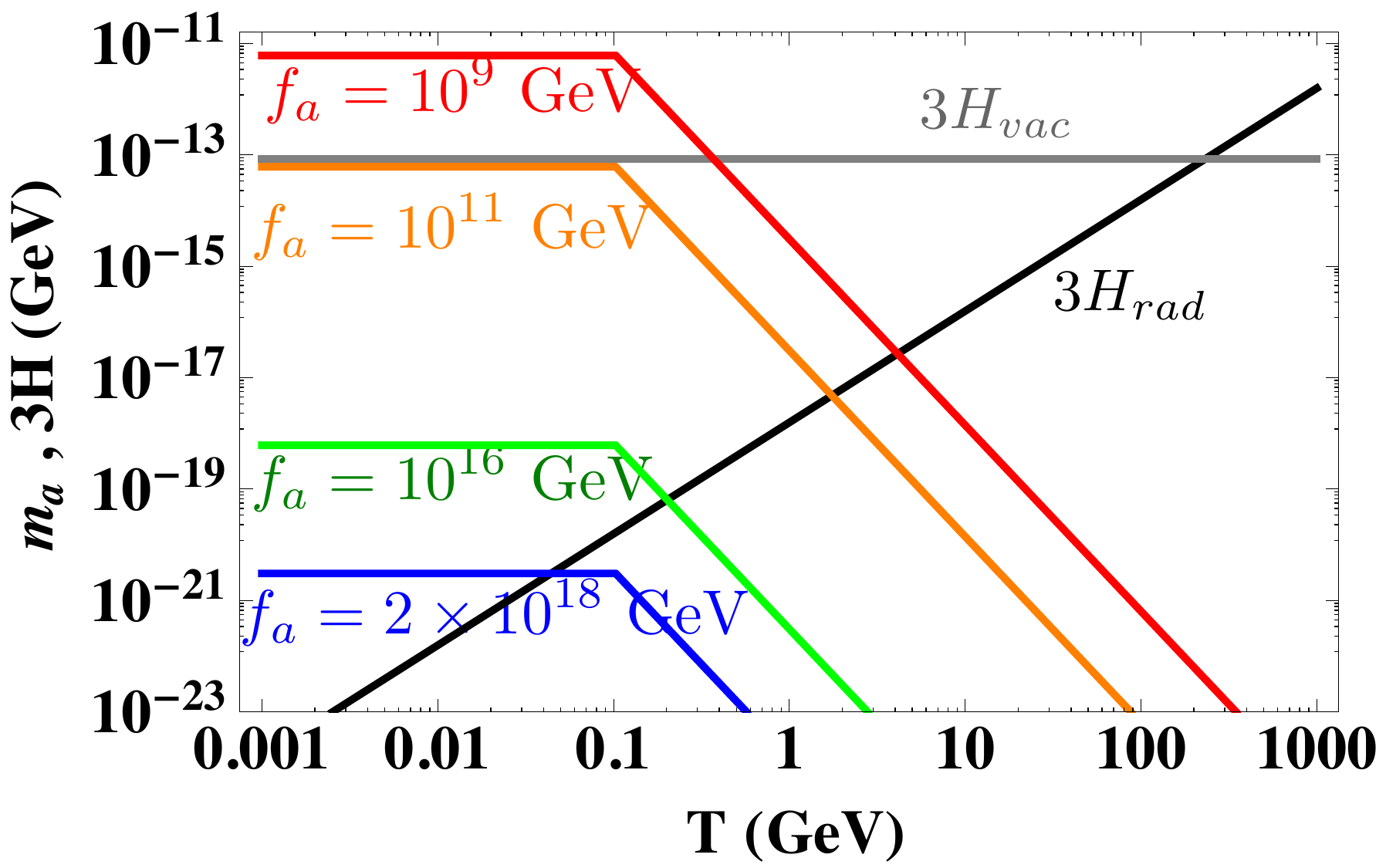}
\caption{\small Comparison between the axion oscillation rate $m_a$ for various PQ scales $f_a$ with the expansion rate of the universe in the radiation era and the supercooling era. Axion oscillations start when $3H\lesssim m_a$. }
\label{fig:oscillations} 
\end{center}
\end{figure}
%

\section{Dark matter relic abundance of the QCD axion}
\label{sec:DM}

So far,  we have not imposed any constraint from dark matter (DM) considerations.
 The supercooled EWPT being followed by reheating at EW scale temperatures, the DM cosmological evolution is roughly unchanged compared to the usual QCD axion scenario.

From Eq.~(\ref{eq:freeseoutTbound}), the maximal value of $\Delta V$ compatible with the cold baryogenesis scenario is $\Delta V\sim \pi^2 g_* (130)^4/30\sim (312 \mbox{ GeV})^4$,
corresponding to the Hubble rate  $H\sim 2.8 \times 10^{-14}$ GeV, which is the typical value of the expansion rate during supercooling due to vacuum energy in the EW sector.
This value is plotted as the horizontal line in Figure \ref{fig:oscillations} where we compare the sizes of the Hubble rate in the radiation-dominated era with the axion oscillation rate $m_a$. For $f_a \gtrsim 7\times 10^{10} $ GeV, oscillations can start only after reheating and DM predictions are the same as usual.
For $f_a \lesssim 7\times 10^{10} $ GeV, oscillations may start in the supercooling stage before the EWPT, say at $T_1$. They stop during reheating following the EWPT. At this point, the axion angle is frozen to $\bar{\Theta}_{EWPT}$. Oscillations start again when the temperature goes down another time below $T_i\sim 1$ GeV but with a smaller initial $\bar{\Theta}_i$. According to Eq.~(\ref{eq:thetaofT}), the corresponding energy density is thus diluted by a factor $[m_a(T_1)/m_a(T_{EWPT})](T_{EWPT}/T_1)^3$ compared to the standard cosmological scenario. To get the correct DM abundance, we may thus have to push $f_a$ to larger values, bringing us back to the first case in which oscillations start after the EWPT and the 
DM from energy stored in axion oscillations follows the standard prediction.

There is one interesting difference though, which concerns 
the contribution to DM from axion particles produced by the decay of strings. The estimate of this contribution suffers from uncertainties, it is typically assumed to be of the same order as the contribution from the axion field misalignment mechanism but may be larger \cite{Hiramatsu:2010yu}. 
In our scenario, such contribution would be reduced by at least $\sim \log (\Lambda_{EW}/\Lambda_{QCD}) \sim7$ e-folds of inflation associated with the supercooling stage.
Together with the baryon asymmetry, this is the only other relevant observable consequence of our supercooling stage.
Our peculiar cosmological scenario has no impact on Big Bang Nucleosynthesis since the universe is reheated well above the MeV scale.

\section{Conclusion}
\label{sec:conclusion}

We have shown that the QCD axion could play a key role in providing the new source of CP violation in baryogenesis, therefore  linking the origin of dark matter to that of the matter antimatter asymmetry of the universe.
This CP-violating source is proportional to the axion mass squared and is therefore highly suppressed before the QCD phase transition.
Baryogenesis through the axion can be achieved provided that the EWPT is delayed down to sub-GeV temperatures due to a coupling between the Higgs field  and an EW scale dilaton.
The nearly conformal dynamics which has been advocated to protect the EW scale 
naturally  provides the condition for Higgs quenching as needed in the framework of cold baryogenesis.
In terms of the QCD angle $\bar{\Theta}=a/f_a$, the produced baryon asymmetry scales as, for $T_{EWPT} \lesssim \Lambda_{QCD}$, 
\be
\frac{n_B}{s} \sim 10^{-8} \left(\frac{T_{eff}}{T_{reh}}\right)^3  \left. \sin \bar{\Theta} \right|_{EWPT} 
\ee
where $T_{eff}$  measures  the effective temperature of Chern--Simons transitions during the quench and $T_{reh}\sim {\cal O}(100)$ GeV is the reheat temperature. 
If axion oscillations started early in a long supercooling stage, $\sin \bar{\Theta}_{EWPT}$ may be significantly smaller than 1. Despite this, we can still have a large baryon asymmetry for 
$T_{eff}/T_{reh} \gtrsim 5$ (see Fig.~\ref{fig:bau}). This is precisely what lattice simulations of cold baryogenesis predict, $T_{eff}/T_{reh} \sim 30-40$  \cite{Tranberg:2006dg}.

The possibility that the axion could be responsible for the matter antimatter asymmetry of the universe had been discarded back in 1992 in Ref.~\cite{Kuzmin:1992up}, while the cold baryognesis proposal was not yet known.
In Ref.~\cite{Kuzmin:1992up}, which was carried out  in the context of standard EW baryogenesis, $T_{eff}$ was taken to be around $\Lambda_{QCD}$, the temperature at which the axion mass is unsuppressed. Therefore,
 there was no way to get a sufficiently large baryon asymmetry since the reheat temperature after the EWPT has to be around the EW scale.
The key point we have stressed here is that in the context of cold baryogenesis, the effective temperature characterizing baryon number violation may be significantly higher than the actual temperature of the universe. Therefore, even if the EWPT takes place at $T\sim \Lambda_{QCD}$ in order for the strong CP violation to be maximal, we can have $T_{eff} \sim {\cal O}(100) $ GeV, as shown by extensive numerical simulations of cold baryogenesis. As a result, a reheat temperature of order $\cal{O}$(100) GeV as expected in models where the dilaton mass is  $ \sim {\cal O}(100) $ GeV is still compatible with a sufficiently large baryon asymmetry. An important constraint is that the reheat temperature after the dilaton gets its VEV and induces the EWPT should not exceed the freese-out temperature $\sim$ 130 GeV, in which case the baryon asymmetry would be washed out.
This results in a bound on the dilaton mass of the order ${\cal O}(100)$ GeV (see Fig.~\ref{fig:Treheat}) and is therefore a testable scenario at the LHC. A dilaton lighter than the Higgs is not excluded experimentally.

In summary, an EW scale dilaton provides a new route for departure from thermal equilibrium
 and can drive a parametric amplification of baryon number violation at low temperatures. This opens the possibility that the strong CP violation via the QCD axion could be responsible for the baryon asymmetry of the universe during a delayed EWPT.
Although the cosmological history in our scenario is non-standard between temperatures of order  100 GeV down to sub-GeV, there are essentially no constraints 
 on this supercooling stage preceding the EWPT as it is followed by reheating and therefore has no impact on Big Bang Nucleosynthesis. Interestingly, this period of supercooling can dilute the axion particles produced by string decays but does not modify the usual predictions from the axion oscillations which start only after reheating.  For the axion to be dark matter in our framework, we therefore typically predict the Peccei-Quinn scale to be in the same range as usual. It would be interesting to refine these dark matter predictions which depend on the details of the reheating process.
 
 The experimental tests of this scenario are  of three very different types:
the usual QCD axion searches,
  LHC searches for an additional dilaton-like field coupled to the Higgs, and a stochastic milli-Hertz gravity wave background  detectable by eLISA as a signature of the delayed EW phase transition \cite{Konstandin:2011dr}.


\section*{Acknowledgments}

I thank Thomas Hambye, Thomas Konstandin, Javier Redondo, Alex Pomarol, Oriol Pujolas and Christophe Grojean for useful discussions.
This work is supported by the Spanish Ministry MICINN under contract FPA2010Ð17747 and by the Generalitat de Catalunya grant 2014Ð SGRÐ1450.



\begin{thebibliography}{99}

\bibitem{Ade:2013zuv} 
  P.~A.~R.~Ade {\it et al.}  [Planck Collaboration],
  ``Planck 2013 results. XVI. Cosmological parameters,''
  arXiv:1303.5076 [astro-ph.CO].

\bibitem{Rummukainen:1998as} 
  K.~Rummukainen, M.~Tsypin, K.~Kajantie, M.~Laine and M.~E.~Shaposhnikov,
  ``The Universality class of the electroweak theory,''
  Nucl.\ Phys.\ B {\bf 532}, 283 (1998)
  [hep-lat/9805013].

\bibitem{Gavela:1993ts}
  M.~B.~Gavela, P.~Hernandez, J.~Orloff and O.~Pene,
  ``Standard model CP violation and baryon asymmetry,''
  Mod.\ Phys.\ Lett.\ A {\bf 9} (1994) 795
  [hep-ph/9312215, hep-ph/9312215].

\bibitem{Konstandin:2003dx}
  T.~Konstandin, T.~Prokopec and M.~G.~Schmidt,
  ``Axial currents from CKM matrix CP violation and electroweak baryogenesis,''
  Nucl.\ Phys.\ B {\bf 679} (2004) 246
  [hep-ph/0309291].

\bibitem{Tranberg:2009de}
  A.~Tranberg, A.~Hernandez, T.~Konstandin and M.~G.~Schmidt,
  ``Cold electroweak baryogenesis with Standard Model CP violation,''
  Phys.\ Lett.\ B {\bf 690} (2010) 207
  [arXiv:0909.4199 [hep-ph]].

\bibitem{Brauner:2012gu}
  T.~Brauner, O.~Taanila, A.~Tranberg and A.~Vuorinen,
  ``Computing the temperature dependence of effective CP violation in the standard model,''
  JHEP {\bf 1211} (2012) 076
  [arXiv:1208.5609 [hep-ph]].

\bibitem{Davidson:2008bu} 
  S.~Davidson, E.~Nardi and Y.~Nir,
  ``Leptogenesis,''
  Phys.\ Rept.\  {\bf 466}, 105 (2008)
  [arXiv:0802.2962 [hep-ph]].
   
\bibitem{Morrissey:2012db} 
  D.~E.~Morrissey and M.~J.~Ramsey-Musolf,
  ``Electroweak baryogenesis,''
ÊÊNew J.\ Phys.\  {\bf 14}, 125003 (2012)
ÊÊ[arXiv:1206.2942 [hep-ph]].
ÊÊ

\bibitem{Konstandin:2013caa} 
  T.~Konstandin,
  ``Quantum Transport and Electroweak Baryogenesis,''
ÊÊPhys.\ Usp.\  {\bf 56}, 747 (2013)
ÊÊ[Usp.\ Fiz.\ Nauk {\bf 183}, 785 (2013)]
ÊÊ[arXiv:1302.6713 [hep-ph]].
ÊÊ


\bibitem{Servant:2013uwa} 
  G.~Servant and S.~Tulin,
  ``Baryogenesis and Dark Matter through a Higgs Asymmetry,''
  Phys.\ Rev.\ Lett.\  {\bf 111}, no. 15, 151601 (2013)
  [arXiv:1304.3464 [hep-ph]].
  
\bibitem{Davidson:2013psa} 
  S.~Davidson, R.~Gonz‡lez Felipe, H.~Ser™dio and J.~P.~Silva,
  ``Baryogenesis through split Higgsogenesis,''
  JHEP {\bf 1311}, 100 (2013)
  [arXiv:1307.6218 [hep-ph]].

\bibitem{Baker:2006ts} 
  C.~A.~Baker, D.~D.~Doyle, P.~Geltenbort, K.~Green, M.~G.~D.~van der Grinten, P.~G.~Harris, P.~Iaydjiev and S.~N.~Ivanov {\it et al.},
  ``An Improved experimental limit on the electric dipole moment of the neutron,''
  Phys.\ Rev.\ Lett.\  {\bf 97}, 131801 (2006)
  [hep-ex/0602020].

\bibitem{Peccei:2006as} 
  R.~D.~Peccei,
  ``The Strong CP problem and axions,''
ÊÊLect.\ Notes Phys.\  {\bf 741}, 3 (2008)
ÊÊ[hep-ph/0607268].
ÊÊ

\bibitem{'tHooft:1986nc} 
  G.~'t Hooft,
  ``How Instantons Solve the U(1) Problem,''
  Phys.\ Rept.\  {\bf 142}, 357 (1986).

\bibitem{Dias:2014osa} 
  A.~G.~Dias, A.~C.~B.~Machado, C.~C.~Nishi, A.~Ringwald and P.~Vaudrevange,
  ``The Quest for an Intermediate-Scale Accidental Axion and Further ALPs,''
  JHEP {\bf 1406}, 037 (2014)
  [arXiv:1403.5760 [hep-ph]].


\bibitem{Kuzmin:1992up}
  V.~A.~Kuzmin, M.~E.~Shaposhnikov and I.~I.~Tkachev,
  ``Strong CP violation, electroweak baryogenesis, and axionic dark matter,''
  Phys.\ Rev.\ D {\bf 45} (1992) 466.


\bibitem{GarciaBellido:1999sv}
  J.~Garcia-Bellido, D.~Y.~Grigoriev, A.~Kusenko and M.~E.~Shaposhnikov,
  ``Non-equilibrium electroweak baryogenesis from preheating after
  inflation,''
  Phys.\ Rev.\  D {\bf 60}, 123504 (1999)
  [arXiv:hep-ph/9902449].

\bibitem{Krauss:1999ng} 
  L.~M.~Krauss and M.~Trodden,
  ``Baryogenesis below the electroweak scale,''
  Phys.\ Rev.\ Lett.\  {\bf 83}, 1502 (1999)
  [hep-ph/9902420].


\bibitem{Copeland:2001qw}
  E.~J.~Copeland, D.~Lyth, A.~Rajantie and M.~Trodden,
  ``Hybrid inflation and baryogenesis at the TeV scale,''
  Phys.\ Rev.\  D {\bf 64}, 043506 (2001)
  [arXiv:hep-ph/0103231].

\bibitem{Cornwall:2001hq}
  J.~M.~Cornwall, D.~Grigoriev, A.~Kusenko,
  ``Resonant amplification of electroweak baryogenesis at preheating,''
  Phys.\ Rev.\  {\bf D64 } (2001)  123518.
  [hep-ph/0106127].

\bibitem{Smit:2002sn}
  J.~Smit and A.~Tranberg,
  ``Chern-Simons number asymmetry from CP-violation during tachyonic
  preheating,''
  arXiv:hep-ph/0210348.

\bibitem{GarciaBellido:2003wd}
  J.~Garcia-Bellido, M.~Garcia-Perez and A.~Gonzalez-Arroyo,
  ``Chern-Simons production during preheating in hybrid inflation models,''
  Phys.\ Rev.\  D {\bf 69}, 023504 (2004)
  [arXiv:hep-ph/0304285].

\bibitem{Tranberg:2003gi}
  A.~Tranberg and J.~Smit,
  ``Baryon asymmetry from electroweak tachyonic preheating,''
  JHEP {\bf 0311}, 016 (2003)
  [arXiv:hep-ph/0310342].

\bibitem{vanTent:2004rc}
  B.~J.~W.~van Tent, J.~Smit and A.~Tranberg,
  ``Electroweak-scale inflation, inflaton-Higgs mixing and the scalar  spectral
  index,''
  JCAP {\bf 0407} (2004) 003
  [arXiv:hep-ph/0404128].

\bibitem{vanderMeulen:2005sp}
  M.~van der Meulen, D.~Sexty, J.~Smit and A.~Tranberg,
  ``Chern-Simons and winding number in a tachyonic electroweak transition,''
  JHEP {\bf 0602} (2006) 029
  [arXiv:hep-ph/0511080].

\bibitem{Tranberg:2006dg}
  A.~Tranberg, J.~Smit and M.~Hindmarsh,
  ``Simulations of Cold Electroweak Baryogenesis: Finite time quenches,''
  JHEP {\bf 0701} (2007) 034
  [arXiv:hep-ph/0610096].

\bibitem{Enqvist:2010fd}
  K.~Enqvist, P.~Stephens, O.~Taanila and A.~Tranberg,
  ``Fast Electroweak Symmetry Breaking and Cold Electroweak Baryogenesis,''
  arXiv:1005.0752 [astro-ph.CO].


\bibitem{D'Onofrio:2014kta} 
  M.~D'Onofrio, K.~Rummukainen and A.~Tranberg,
  ``The Sphaleron Rate in the Minimal Standard Model,''
  arXiv:1404.3565 [hep-ph].

\bibitem{Lue:1996pr} 
  A.~Lue, K.~Rajagopal and M.~Trodden,
  ``Semianalytical approaches to local electroweak baryogenesis,''
  Phys.\ Rev.\ D {\bf 56}, 1250 (1997)
  [hep-ph/9612282].

\bibitem{Baron:2013eja}
  J.~Baron {\it et al.}  [ACME Collaboration],
  ``Order of Magnitude Smaller Limit on the Electric Dipole Moment of the Electron,''
  Science {\bf 343} (2014) 6168,  269
  [arXiv:1310.7534 [physics.atom-ph]].

\bibitem{Konstandin:2011ds}
  T.~Konstandin and G.~Servant,
  ``Natural Cold Baryogenesis from Strongly Interacting Electroweak Symmetry Breaking,''
  JCAP {\bf 1107} (2011) 024
  [arXiv:1104.4793 [hep-ph]].

\bibitem{Wantz:2009it} 
  O.~Wantz and E.~P.~S.~Shellard,
  ``Axion Cosmology Revisited,''
  Phys.\ Rev.\ D {\bf 82}, 123508 (2010)
  [arXiv:0910.1066 [astro-ph.CO]].

\bibitem{Buchmuller:1990ds} 
  W.~Buchmuller and D.~Wyler,
  ``The Effect of dilatons on the electroweak phase transition,''
  Phys.\ Lett.\ B {\bf 249}, 281 (1990).


\bibitem{Konstandin:2011dr}
  T.~Konstandin and G.~Servant,
  ``Cosmological Consequences of Nearly Conformal Dynamics at the TeV scale,''
  JCAP {\bf 1112} (2011) 009
  [arXiv:1104.4791 [hep-ph]].


\bibitem{Witten:1980ez} 
  E.~Witten,
  ``Cosmological Consequences of a Light Higgs Boson,''
  Nucl.\ Phys.\ B {\bf 177}, 477 (1981).
%
ÊÊ
ÊÊ
        
   
\bibitem{Asaka:2001ez} 
  T.~Asaka, W.~Buchmuller and L.~Covi,
  ``False vacuum decay after inflation,''
ÊÊPhys.\ Lett.\ B {\bf 510}, 271 (2001)
ÊÊ[hep-ph/0104037].
ÊÊ
  

\bibitem{Copeland:2002ku} 
  E.~J.~Copeland, S.~Pascoli and A.~Rajantie,
  ``Dynamics of tachyonic preheating after hybrid inflation,''
ÊÊPhys.\ Rev.\ D {\bf 65}, 103517 (2002)
ÊÊ[hep-ph/0202031].
ÊÊ
  
\bibitem{GarciaBellido:2002aj} 
  J.~Garcia-Bellido, M.~Garcia Perez and A.~Gonzalez-Arroyo,
  ``Symmetry breaking and false vacuum decay after hybrid inflation,''
ÊÊPhys.\ Rev.\ D {\bf 67}, 103501 (2003)
ÊÊ[hep-ph/0208228].
ÊÊ
  
\bibitem{Smit:2002yg} 
  J.~Smit and A.~Tranberg,
  ``Chern-Simons number asymmetry from CP violation at electroweak tachyonic preheating,''
ÊÊJHEP {\bf 0212}, 020 (2002)
ÊÊ[hep-ph/0211243].
ÊÊ


  
\bibitem{Borsanyi:2003ib} 
  S.~Borsanyi, A.~Patkos and D.~Sexty,
  ``Nonequilibrium Goldstone phenomenon in tachyonic preheating,''
ÊÊPhys.\ Rev.\ D {\bf 68}, 063512 (2003)
ÊÊ[hep-ph/0303147].
ÊÊ
 
\bibitem{Randall:1999ee} 
  L.~Randall and R.~Sundrum,
  Phys.\ Rev.\ Lett.\  {\bf 83}, 3370 (1999)
  [hep-ph/9905221].

\bibitem{Goldberger:1999uk} 
  W.~D.~Goldberger and M.~B.~Wise,
  ``Modulus stabilization with bulk fields,''
  Phys.\ Rev.\ Lett.\  {\bf 83}, 4922 (1999)
  [hep-ph/9907447].

\bibitem{Rattazzi:2000hs}
  R.~Rattazzi and A.~Zaffaroni,
  ``Comments on the holographic picture of the Randall-Sundrum model,''
  JHEP {\bf 0104} (2001) 021
  [hep-th/0012248].

\bibitem{Garriga:2002vf} 
  J.~Garriga and A.~Pomarol,
  ``A Stable hierarchy from Casimir forces and the holographic interpretation,''
  Phys.\ Lett.\ B {\bf 560}, 91 (2003)
  [hep-th/0212227].

\bibitem{Creminelli:2001th}
  P.~Creminelli, A.~Nicolis and R.~Rattazzi,
  ``Holography and the electroweak phase transition,''
  JHEP {\bf 0203} (2002) 051
  [arXiv:hep-th/0107141].

\bibitem{Randall:2006py}
  L.~Randall and G.~Servant,
  ``Gravitational Waves from Warped Spacetime,''
  JHEP {\bf 0705} (2007) 054
  [arXiv:hep-ph/0607158].

\bibitem{Konstandin:2010cd}
  T.~Konstandin, G.~Nardini and M.~Quiros,
  ``Gravitational Backreaction Effects on the Holographic Phase Transition,''
  arXiv:1007.1468 [hep-ph].

   
\bibitem{Coradeschi:2013gda}
  F.~Coradeschi, P.~Lodone, D.~Pappadopulo, R.~Rattazzi and L.~Vitale,
  ``A naturally light dilaton,''
  JHEP {\bf 1311} (2013) 057
  [arXiv:1306.4601 [hep-th]].

\bibitem{Bellazzini:2013fga} 
  B.~Bellazzini, C.~Csaki, J.~Hubisz, J.~Serra and J.~Terning,
  ``A Naturally Light Dilaton and a Small Cosmological Constant,''
  arXiv:1305.3919 [hep-th].


\bibitem{Megias:2014iwa}
  E.~Megias and O.~Pujolas,
  ``Naturally light dilatons from nearly marginal deformations,''
  arXiv:1401.4998 [hep-th].
  
\bibitem{Chacko:2013dra} 
  Z.~Chacko, R.~K.~Mishra and D.~Stolarski,
  ``Dynamics of a Stabilized Radion and Duality,''
  JHEP {\bf 1309}, 121 (2013)
  [arXiv:1304.1795 [hep-ph]].
  
   
\bibitem{Coleppa:2011zx} 
  B.~Coleppa, T.~Gregoire and H.~E.~Logan,
  ``Dilaton constraints and LHC prospects,''
  Phys.\ Rev.\ D {\bf 85}, 055001 (2012)
  [arXiv:1111.3276 [hep-ph]].
   
\bibitem{Barger:2011hu} 
  V.~Barger, M.~Ishida and W.~-Y.~Keung,
  ``Differentiating the Higgs boson from the dilaton and the radion at hadron colliders,''
  Phys.\ Rev.\ Lett.\  {\bf 108}, 101802 (2012)
  [arXiv:1111.4473 [hep-ph]].
   
\bibitem{Campbell:2011iw} 
  B.~A.~Campbell, J.~Ellis and K.~A.~Olive,
  ``Phenomenology and Cosmology of an Electroweak Pseudo-Dilaton and Electroweak Baryons,''
  JHEP {\bf 1203}, 026 (2012)
  [arXiv:1111.4495 [hep-ph]].
   
\bibitem{deSandes:2011zs} 
  H.~de Sandes and R.~Rosenfeld,
  ``Radion-Higgs mixing effects on bounds from LHC Higgs Searches,''
  Phys.\ Rev.\ D {\bf 85}, 053003 (2012)
  [arXiv:1111.2006 [hep-ph]].
   
\bibitem{Goldberger:2008zz} 
  W.~D.~Goldberger, B.~Grinstein and W.~Skiba,
  ``Distinguishing the Higgs boson from the dilaton at the Large Hadron Collider,''
  Phys.\ Rev.\ Lett.\  {\bf 100}, 111802 (2008)
  [arXiv:0708.1463 [hep-ph]].
  
\bibitem{Bellazzini:2012vz} 
  B.~Bellazzini, C.~Csaki, J.~Hubisz, J.~Serra and J.~Terning,
  ``A Higgslike Dilaton,''
  Eur.\ Phys.\ J.\ C {\bf 73}, 2333 (2013)
  [arXiv:1209.3299 [hep-ph]].
  

\bibitem{Chacko:2012vm} 
  Z.~Chacko, R.~Franceschini and R.~K.~Mishra,
  ``Resonance at 125 GeV: Higgs or Dilaton/Radion?,''
  JHEP {\bf 1304}, 015 (2013)
  [arXiv:1209.3259 [hep-ph]].

   
\bibitem{Desai:2013pga} 
  N.~Desai, U.~Maitra and B.~Mukhopadhyaya,
  ``An updated analysis of radion-higgs mixing in the light of LHC data,''
  JHEP {\bf 1310}, 093 (2013)
  [arXiv:1307.3765 [hep-ph]].
   
\bibitem{Cao:2013cfa} 
  J.~Cao, Y.~He, P.~Wu, M.~Zhang and J.~Zhu,
  ``Higgs Phenomenology in the Minimal Dilaton Model after Run I of the LHC,''
  JHEP {\bf 1401}, 150 (2014)
  [arXiv:1311.6661 [hep-ph]].
  
\bibitem{Jung:2014zga} 
  D.~-W.~Jung and P.~Ko,
  ``Higgs-dilaton(radion) system confronting the LHC Higgs data,''
  Phys.\ Lett.\ B {\bf 732}, 364 (2014)
  [arXiv:1401.5586 [hep-ph]].

\bibitem{Hiramatsu:2010yu} 
  T.~Hiramatsu, M.~Kawasaki, T.~Sekiguchi, M.~Yamaguchi and J.~'i.~Yokoyama,
  ``Improved estimation of radiated axions from cosmological axionic strings,''
  Phys.\ Rev.\ D {\bf 83}, 123531 (2011)
  [arXiv:1012.5502 [hep-ph]].


   
\end{thebibliography}
\end{document}